# Polarization-Dependent Loss of Optical Connectors Measured with High Accuracy (<0.004 dB) after Cancelation of Polarimetric Errors

Reinhold Noé[(1,2)], Benjamin Koch[(1,2)]

(1) Paderborn University, EIM-E, Warburger Str. 100, D-33098 Paderborn, Germany, e-mail: noe@upb.de
(2) Novoptel GmbH, Helmerner Weg 2, D-33100 Paderborn, Germany, E-mail: info@novoptel.com

*Abstract*—State-of-the-art polarimeter calibration is reviewed. Producing many quasi-random polarization states and moving/bending a fiber without changing power allows finding a polarimeter calibration where the degree-of-polarization reaches unity and parasitic polarization-dependent loss is small.

Using a polarization scrambler/transformer and a polarimeter a device-under-test can be characterized. Its Mueller matrix can be decomposed into a product of a nondepolarizing Mueller-Jones matrix times a purely depolarizing Mueller matrix.

Test polarizations may drift over time. With help of an optical switch the reference device can be measured against an internal reference path. Later, with possibly different test polarizations, the actual device-under-test is measured against the internal reference. Polarization drift and need for repeated reference device measurement are thus overcome.

When a patchcord is inserted, connector PDL can be measured, provided that errors are calibrated away, again by fiber moving/bending.

Experimentally we have measured PDL with errors <0.004 dB. This easily suffices to measure connector PDL, which is demonstrated. PDL >60 dB was measured when the device under test was a good polarizer. A 20 Mrad/s polarization scrambler with $LiNbO_3$ device generates the test polarizations. The polarimeter can sample at 100 MHz and can store 64M Stokes vectors. During laser frequency scans Mueller matrices can be measured in time intervals as short as 5 μs.

## I. Introduction

Polarization-dependent loss (PDL) is an important property of fiberoptic components, from polarizers (very high PDL) to connectors (very low PDL). Various measurement methods have been reported.

Most of them are based on pure intensity measurements [1-5] behind a device-under-test (DUT). This requires components with low PDL, namely photodetectors and connectors. Test polarizations are generated by a polarization transformer or scrambler which itself has PDL or is mechanic and slow.

With an additional polarimeter it is possible to measure the complete Mueller matrix of the DUT [6]. This is fast and allows measurement even of highest PDL >60 dB [7].

In [7] the above methods are experimentally compared. For the measurement of lowest PDL it is not very clear which method is most accurate.

However, for connector testing, fast measurement of PDL <0.01 dB is needed. This doesn't seem to have been reported.

The displayed degree-of-polarization (DOP) is a very useful measure for the accuracy of a polarimeter calibration. This was found and exploited by J. Rasmussen [8]. Many different polarization states with constant (usually unity) DOP are generated. The polarimeter calibration matrix is optimized until the measured DOP is constant. Without the iterative optimization, the usage of many different, equidistributed polarization states for polarimeter calibration has also been described in [9]. E. Krause and A. Bandemer [10] have described the iterative improvement of the polarimeter calibration matrix for constant DOP, very similar to [8].

In this paper we describe how a polarimeter can be accurately calibrated for Mueller matrix measurement. In the process, various polarimetric errors are canceled. Using obtained calibration data, Mueller matrices are measured with a PDL uncertainty of <0.004 dB. This easily allows measuring PDL of exemplary connectors of several types, with higher accuracy than previously possible. Accuracy is <0.03 dB / <0.01 dB / <0.005 dB / <0.004 dB for 5 μs / 100 μs / 10 ms / 100 ms measurement time. A fast $LiNbO_3$ polarization transformer and a polarimeter and exact PDL and drift calibration are decisive for this.

## II. Basic Polarimeter Calibration

We describe basic polarimeter calibration similar to [8, 9]. It will later be expanded to the calibration of the whole setup for Mueller matrix measurement.

A polarimeter with beamsplitters/polarizers/waveplates and 4 photodiodes generally yields a 4×1 photocurrent vector **I**. More generally we write

$$\mathbf{X} = \mathbf{FS}. \tag{1}$$

For the time being, $\mathbf{X} = \mathbf{I}$ and $\mathbf{I} = \mathbf{FS}$. Here it is assumed that dark currents or offsets are already subtracted. $\mathbf{S} = [S_0, S_1, S_2, S_3]^T$ is the 4×1 Stokes vector and $\mathbf{F}$ is a 4×4 matrix. The whole calculus can easily be extended to more than 4 photodiodes. One can calculate the inverse $\mathbf{S} = \mathbf{F}^{-1}\mathbf{I}$. Polarimeter calibration consists in finding matrix $\mathbf{F}^{-1}$.

Polarization states with Stokes vectors $\mathbf{S}_i$ ($i = 1...n; n \geq 4$) are generated by a polarization scrambler. This way we get 4×n matrices

$$\mathbf{I}_a = \mathbf{FS}_a \qquad \mathbf{S}_a = [\mathbf{S}_1, \mathbf{S}_2, ..., \mathbf{S}_n]. \tag{2}$$

Let us assume equidistributed polarizations. In that case the average is unpolarized light with photocurrent and Stokes vectors

$$\langle \mathbf{I}_a \rangle = \langle \mathbf{FS}_a \rangle = \mathbf{F}\langle \mathbf{S}_a \rangle, \tag{3}$$

$$\langle \mathbf{S}_a \rangle = \begin{bmatrix} 1 & 0 & 0 & 0 \end{bmatrix}^T. \tag{4}$$

One specific polarization state is set or defined as horizontal polarization. The measurement yields

$$\mathbf{I}_b = \mathbf{FS}_b \qquad \mathbf{S}_b = \begin{bmatrix} 1 & 1 & 0 & 0 \end{bmatrix}^T. $$

Another polarization is set or defined to be linear with unknown orientation angle $0 < \vartheta < \pi/2$, preferably near $\pi/4$. We measure

$$\mathbf{I}_c = \mathbf{FS}_c \qquad \mathbf{S}_c = \begin{bmatrix} 1 & \cos 2\vartheta & \sin 2\vartheta & 0 \end{bmatrix}^T. \tag{5}$$

The covariance matrix of the $n$ intensity vectors is

$$\langle \mathbf{II}^T \rangle = \mathbf{F}\langle \mathbf{S}_a \mathbf{S}_a^T \rangle \mathbf{F}^T, \tag{6}$$

$$\langle \mathbf{S}_a \mathbf{S}_a^T \rangle = \begin{bmatrix} 1 & 1 & 0 & 0 \\ 0 & 1/3 & 0 & 0 \\ 0 & 0 & 1/3 & 0 \\ 0 & 0 & 0 & 1/3 \end{bmatrix}. \tag{7}$$

The latter expression $\langle \mathbf{S}_a \mathbf{S}_a^T \rangle$ follows from equidistributed polarizations. We diagonalize $\langle \mathbf{II}^T \rangle$ as

$$\langle \mathbf{II}^T \rangle = \mathbf{A\Lambda}_a \mathbf{A}^T = \mathbf{BB}^T = \mathbf{BCC}^T\mathbf{B}^T$$
$$\mathbf{B} = \mathbf{A}\sqrt{\mathbf{\Lambda}_a} \qquad \mathbf{AA}^T = \mathbf{1} \qquad \mathbf{CC}^T = \mathbf{1} \tag{8}$$

with eigenvalue matrix $\mathbf{\Lambda}_a$ and orthogonal eigenvector matrix A. The symmetric covariance matrix does not allow a unique decomposition because it remains unchanged when any orthogonal matrix $\mathbf{C}$ times its transpose (= inverse) is inserted. $\mathbf{C}$ can be chosen to achieve the desired polarimeter calibration, freely except that $\mathbf{C}$ must be made orthogonal. We equate the decomposed halves of the two $\langle \mathbf{II}^T \rangle$ expressions (6), (8) and solve for

$$\mathbf{F} = \mathbf{BC}\begin{bmatrix} 1 & 1 & 0 & 0 \\ 0 & \sqrt{3} & 0 & 0 \\ 0 & 0 & \sqrt{3} & 0 \\ 0 & 0 & 0 & \sqrt{3} \end{bmatrix}. \tag{9}$$

In the following column indices 0...3 are used for square matrices such as $\mathbf{F}$ and $\mathbf{C}$, like the indices of the Stokes parameters. Unpolarized light delivers

$$\mathbf{F}_0 = \langle \mathbf{I}_a \rangle \qquad \mathbf{C}_0 = \mathbf{B}^{-1}\mathbf{F}_0. \tag{10}$$

Horizontal polarization yields

$$\mathbf{F}_0 + \mathbf{F}_1 = \mathbf{I}_b \qquad \mathbf{C}_1 = \mathbf{B}^{-1}(\mathbf{I}_b - \mathbf{F}_0)/\sqrt{3}. \tag{11}$$

Practically, (4), (7) may not be exactly fulfilled. This means $\mathbf{C}_0$, $\mathbf{C}_1$ are not exactly orthogonal. One can orthogonalize them by appropriately changing one or both.

For linear polarization it holds

$$\mathbf{F}_0 + \mathbf{F}_1 \cos 2\vartheta + \mathbf{F}_2 \sin 2\vartheta = \mathbf{I}_c$$
$$\hat{\mathbf{C}}_2 = \mathbf{C}_1 \cos 2\vartheta + \mathbf{C}_2 \sin 2\vartheta = \mathbf{B}^{-1}(\mathbf{I}_c - \mathbf{F}_0)/\sqrt{3}. \tag{12}$$

Under the assumption of orthogonality the auxiliary vector $\hat{\mathbf{C}}_2$ allows calculating $\mathbf{C}_2$,

$$\mathbf{C}_1^T \hat{\mathbf{C}}_2 = |\mathbf{C}_1|^2 \cos 2\vartheta, \tag{13}$$

$$\hat{\mathbf{C}}_2 - \mathbf{C}_1(\mathbf{C}_1^T \hat{\mathbf{C}}_2)/|\mathbf{C}_1|^2 = \mathbf{C}_2 \sin 2\vartheta, \tag{14}$$

$$\mathbf{C}_2 = (\mathbf{C}_2 \sin 2\vartheta)/|\mathbf{C}_2 \sin 2\vartheta|. \tag{15}$$

$\mathbf{C}_3$ is found by searching a normalized vector that is orthogonal to $\mathbf{C}_0$, $\mathbf{C}_1$, $\mathbf{C}_2$. Its sign defines the polarization ellipse handedness. Knowing $\mathbf{C}$, matrix $\mathbf{F}^{-1}$ can now be calculated by inversion of (9). Yet, if (4), (7) are not exactly met, $\mathbf{F}^{-1}$ will normally be inaccurate.

But according to [8, 10] the matrix $\mathbf{F}^{-1}$ can be improved by trying to bring the DOP close to 1 for all test polarizations. One simple implementation of the great idea presented in [8] is the following. An available set $\mathbf{I}$ of many measurements is chosen, in particular $\mathbf{I} = \mathbf{I}_a$. The current $\mathbf{F}$ becomes a starting matrix $\mathbf{F}'_0 = \mathbf{F}$. Starting with $l = 0$ one calculates

$$\mathbf{S}_l = \mathbf{F}'^{-1}_l \mathbf{I}. \tag{16}$$

Each column vector $\mathbf{S}_{l,i}$ of matrix $\mathbf{S}_l$ is modified for the next iteration to have constant DOP, in particular the usually expected DOP = 1:

$$\mathbf{S}_{l+1,i} = S_{0,l,i} \left[ 1 \quad \frac{[S_{1,l,i} \quad S_{2,l,i} \quad S_{3,l,i}]}{\sqrt{S_{1,l,i}^2 + S_{2,l,i}^2 + S_{3,l,i}^2}} \right] \tag{17}$$

Normally one wants to compensate also residual PDL contained in the polarimeter calibration. In that case one expects constant intensity and replaces all $S_{0,l,i}$ by the mean of the calculated $S_{0,l,i}$ over all $i$, or by another constant if the applied optical power is known. With the matrix $\mathbf{S}_{l+1}$ of all modified $\mathbf{S}_{l+1,i}$ vectors one can write

$$\mathbf{I} \stackrel{!}{=} \mathbf{F}'_{l+1} \mathbf{S}_{l+1} \tag{18}$$

which in general is not exactly fulfilled. It is solved for the improved $\mathbf{F}'_{l+1}$, namely by

$$\mathbf{F}'_{l+1} = \left(\mathbf{IS}_{l+1}^T\right)\left(\mathbf{S}_{l+1}\mathbf{S}_{l+1}^T\right)^{-1}. \tag{19}$$

The process (16)–(19) is repeated several times until it has converged. The last $\mathbf{F}' = \lim_{l \to \infty} \mathbf{F}'_{l+1}$ replaces $\mathbf{F}$. Inversion yields the desired improved $\mathbf{F}'^{-1}$.

All the foregoing was not limited to $\mathbf{X} = \mathbf{I}$ being intensity vectors. Instead, $\mathbf{X} = \breve{\mathbf{S}}$ could be the imperfect Stokes vectors $\breve{\mathbf{S}} = \breve{\mathbf{F}}^{-1}\mathbf{I}$ which are output by an imperfectly calibrated polarimeter. This means one can improve an existing polarimeter calibration. Here it makes sense to start with $\mathbf{F}_0' = \mathbf{1}$. The improved output is then

$$\mathbf{S} = \mathbf{F}'^{-1}\breve{\mathbf{S}} = \mathbf{F}'^{-1}\breve{\mathbf{F}}^{-1}\mathbf{I} \quad (20)$$

where $\mathbf{F}'^{-1}\breve{\mathbf{F}}^{-1}$ is the improved calibration matrix.

Likewise, one can modify an existing polarimeter calibration for only a rotation of the Poincaré sphere. In that case $\mathbf{F}'^{-1} = \mathbf{R}$ must be a retarder matrix

$$\mathbf{R} = \begin{bmatrix} 1 & 0 & 0 & 0 \\ 0 & & & \\ 0 & & \mathbf{G} & \\ 0 & & & \end{bmatrix} \quad \mathbf{G}\mathbf{G}^T = \mathbf{1}. \quad (21)$$

The first row vector $\mathbf{G}_1$ of the rotation matrix $\mathbf{G}$ is chosen equal to that measured normalized Stokes vector which shall now become horizontal. To find the second row vector $\mathbf{G}_2$, one first sets a preliminary $\hat{\mathbf{G}}_2$ equal to a measured Stokes vector which shall now become linear polarization with $0 < \vartheta < \pi/2$. One must subtract its components parallel to $\mathbf{G}_1$. $\mathbf{G}_2$ is the normalized version of the subtraction result $\hat{\mathbf{G}}_2 - \mathbf{G}_1\left(\mathbf{G}_1^T\hat{\mathbf{G}}_2\right)/|\mathbf{G}_1|^2$ (where $|\mathbf{G}_1|^2 = 1$ anyway). The third row vector is $\mathbf{G}_3 = \mathbf{G}_1 \times \mathbf{G}_2$.

Residual polarization-dependent loss (PDL), for instance after a connector change at the polarimeter input can also be calibrated away. Here is explained how this can be done without at the same time influencing the DOP. Let us assume

$$\breve{\mathbf{S}} = \mathbf{P}\mathbf{S}, \qquad \mathbf{S} = \mathbf{P}^{-1}\breve{\mathbf{S}}. \quad (22)$$

The incorrectly measured Stokes vector $\breve{\mathbf{S}}$ equals an unknown symmetric PDL matrix $\mathbf{P}$ (23) times the unknown correct Stokes vector $\mathbf{S}$. All matrices $\mathbf{P}$, also with subscripts, are symmetric! $\mathbf{V} = [V_1 \; V_2 \; V_3]^T$ with $V_1^2 + V_2^2 + V_3^2 = 1$ is the normalized Stokes (eigen)vector of the polarization with strongest transmission $T_{\max}$, assuming positive extinction unit $\gamma > 0$. Weakest transmission $T_{\min}$ occurs for input polarization $-\mathbf{V}$. Their quotient is $T_{\max}/T_{\min} = e^{2\gamma}$.

Geometric average transmission is $T_{ga} = \sqrt{T_{\max}T_{\min}}$. The extinction unit is $\gamma$ [5] (called $a_j$ in [11, 12]). In Section VI., experimental results will be given as length $(= |\mathbf{\Gamma}_{dB}| = \frac{20}{\ln 10}\gamma$ = PDL in dB) and elements of the dB-scaled PDL vector

$$\mathbf{\Gamma}_{dB} = \frac{20}{\ln 10}\gamma \cdot [V_1 \; V_2 \; V_3]^T. \quad (24)$$

It is advantageous when PDL devices are cascaded: They can be added exactly if their directions are identical (or opposed). They can be added with good accuracy even if their directions differ; exact expressions are given in [5].

All (input) Stokes vectors $\mathbf{S}_i$ must have constant power, i.e. the uppermost line $\mathbf{S}_0$ of $\mathbf{S}$ has identical constant elements $S_0$, for instance equal to the geometric mean of the measured $\breve{\mathbf{S}}_0$. This means

$$S_0 \cdot [1 \; 1 \; \ldots \; 1] = \mathbf{S}_0 = \mathbf{P}_0^{-1}\breve{\mathbf{S}} \quad (25)$$

where $\mathbf{P}_0^{-1}$ is the uppermost line of $\mathbf{P}^{-1}$. This can be solved for $\mathbf{P}_0^{-1}$, in particular by

$$\mathbf{P}_0^{-1} = \mathbf{S}_0\breve{\mathbf{S}}^T\left(\breve{\mathbf{S}}\breve{\mathbf{S}}^T\right)^{-1}. \quad (26)$$

One may wish to set $T_{ga} = 1$.

Comparison with (23) allows calculating all elements of $\mathbf{F}'^{-1} = \mathbf{P}^{-1}$, in order to compute the correct $\mathbf{S}$ from the incorrect $\breve{\mathbf{S}}$. Residual polarimeter PDL is thereby compensated.

### III. MUELLER-JONES MATRICES, DEPOLARIZATION

The Mueller matrix $\mathbf{M}$ has 16 degrees-of-freedom (DOF). But a device for which a Jones matrix exists has a Mueller matrix with only 7 DOF, the so-called Mueller-Jones matrix $\mathbf{M}_J$ (27). In the $\pm$ signs of (27) the + holds for ellipticity handedness in an $x$-$y$-$t$ coordinate system, the – for the (more

$$\mathbf{P} = T_{ga}\begin{bmatrix} \cosh\gamma & V_1\sinh\gamma & V_2\sinh\gamma & V_3\sinh\gamma \\ V_1\sinh\gamma & 1+V_1^2(\cosh\gamma-1) & V_1V_2(\cosh\gamma-1) & V_1V_3(\cosh\gamma-1) \\ V_2\sinh\gamma & V_1V_2(\cosh\gamma-1) & 1+V_2^2(\cosh\gamma-1) & V_2V_3(\cosh\gamma-1) \\ V_3\sinh\gamma & V_1V_3(\cosh\gamma-1) & V_2V_3(\cosh\gamma-1) & 1+V_3^2(\cosh\gamma-1) \end{bmatrix} \quad \text{partial polarizer matrix (23)}$$

$$\mathbf{M}_J = \frac{1}{2}\begin{bmatrix} |J_{11}|^2+|J_{12}|^2+|J_{21}|^2+|J_{22}|^2 & |J_{11}|^2-|J_{12}|^2+|J_{21}|^2-|J_{22}|^2 & 2\mathrm{Re}\!\begin{pmatrix}J_{11}J_{12}^*\\+J_{21}J_{22}^*\end{pmatrix} & \pm 2\mathrm{Im}\!\begin{pmatrix}-J_{11}J_{12}^*\\-J_{21}J_{22}^*\end{pmatrix} \\ |J_{11}|^2+|J_{12}|^2-|J_{21}|^2-|J_{22}|^2 & |J_{11}|^2-|J_{12}|^2-|J_{21}|^2+|J_{22}|^2 & 2\mathrm{Re}\!\begin{pmatrix}J_{11}J_{12}^*\\-J_{21}J_{22}^*\end{pmatrix} & \pm 2\mathrm{Im}\!\begin{pmatrix}-J_{11}J_{12}^*\\+J_{21}J_{22}^*\end{pmatrix} \\ 2\mathrm{Re}\!\begin{pmatrix}J_{11}J_{21}^*\\+J_{12}J_{22}^*\end{pmatrix} & 2\mathrm{Re}\!\begin{pmatrix}J_{11}J_{21}^*\\-J_{12}J_{22}^*\end{pmatrix} & 2\mathrm{Re}\!\begin{pmatrix}J_{11}J_{22}^*\\+J_{12}J_{21}^*\end{pmatrix} & \pm 2\mathrm{Im}\!\begin{pmatrix}-J_{11}J_{22}^*\\+J_{12}J_{21}^*\end{pmatrix} \\ \pm 2\mathrm{Im}\!\begin{pmatrix}J_{11}J_{21}^*\\+J_{12}J_{22}^*\end{pmatrix} & \pm 2\mathrm{Im}\!\begin{pmatrix}J_{11}J_{21}^*\\-J_{12}J_{22}^*\end{pmatrix} & \pm 2\mathrm{Im}\!\begin{pmatrix}J_{11}J_{22}^*\\+J_{12}J_{21}^*\end{pmatrix} & 2\mathrm{Re}\!\begin{pmatrix}J_{11}J_{22}^*\\-J_{12}J_{21}^*\end{pmatrix} \end{bmatrix} \quad \text{Mueller-Jones matrix} \quad (27)$$

common) x-y-z. $\mathbf{M}_J$ can be calculated from the elements of the Jones matrix without any averaging. $\mathbf{M}_J$ is nondepolarizing. An invertible linear operation $\mathbf{H}$ [13-15] allows transforming it into a matrix $\mathbf{Q}_J = \mathbf{H}(\mathbf{M}_J)$ which equals

$$\mathbf{Q}_J = \mathbf{J}_V \mathbf{J}_V^+ \qquad \mathbf{J}_V = [J_{11} \quad J_{12} \quad J_{21} \quad J_{22}]^T. \qquad (28)$$

Only one of its eigenvalues is non-zero (equal to 2 if the device is lossless) and it belongs to the vector $\mathbf{J}_V$ (which is a not normalized eigenvector of $\mathbf{Q}_J$). Comparison of (27) and (28) concretely yields coefficients of $\mathbf{H}$ and its inverse $\mathbf{H}^{-1}$. Each element of $\mathbf{Q}_J$ is a linear combination of 4 elements of $\mathbf{M}_J$ (and vice versa).

One can condense a Mueller matrix $\mathbf{M}$ into a Mueller-Jones matrix $\mathbf{M}_J$. This makes sense if one knows that $\mathbf{M}$ represents a device for which $\mathbf{M}_J$ exists. In doing so, measurement errors of $\mathbf{M}$ can partly be eliminated. To obtain $\mathbf{M}_J$ one takes the Hermitian matrix $\mathbf{Q} = \mathbf{H}(\mathbf{M})$ and diagonalizes it,

$$\mathbf{Q} = \mathbf{K}\boldsymbol{\Lambda}\mathbf{K}^+. \qquad (29)$$

Then

$$\mathbf{Q}_J = \mathbf{K}_0 \lambda_0 \mathbf{K}_0^+ \qquad (30)$$

is formed where $\mathbf{K}_0$ (proportional to $\mathbf{J}_V$) is the eigenvector corresponding to the maximum eigenvalue $\lambda_0$. This eliminates the depolarizing influence of the other eigenvalues. Finally $\mathbf{M}_J = \mathbf{H}^{-1}(\mathbf{Q}_J)$ is obtained by backtransformation. We write this whole process as

$$\mathbf{M}_J = \mathbf{N}(\mathbf{M}) \qquad (31)$$

where the letter $\mathbf{N}$ means nondepolarizing. Clearly, the matrix $\mathbf{M} - \mathbf{N}(\mathbf{M})$ has 9 DOF.

One may define a mean depolarization [6]

$$\overline{d} = (4/3)\left(\sum_{i=1}^{3} \lambda_i\right) \Big/ \left(\sum_{i=0}^{3} \lambda_i\right). \qquad (32)$$

This is identical with $\overline{d} = (1/3)\sum_{i=1}^{3}(1 - \text{DOP}_i)$ that is obtained for a partial depolarizer with diagonal Mueller matrix $\mathbf{M} = \text{diag}(1, \text{DOP}_1, \text{DOP}_2, \text{DOP}_3)$. $\text{DOP}_i$ indicates how much the Stokes parameter $S_i$ is depolarized.

One may need to factorize a Mueller matrix into a nondepolarizing Mueller-Jones matrix $\mathbf{Z}_J$ with 7 DOF times a depolarizer matrix $\mathbf{D}$ with 9 DOF,

$$\mathbf{M} = \mathbf{Z}_J \mathbf{D} \qquad \mathbf{N}(\mathbf{D}) = D \cdot \mathbf{1} \qquad \mathbf{N}(\mathbf{Z}_J) = \mathbf{Z}_J. \qquad (33)$$

$D$ is a scalar. Clearly, $\mathbf{M}$ may be depolarizing. The nondepolarizing content of the depolarizer matrix $\mathbf{D}$ is equal or proportional to the unity matrix $\mathbf{1}$. Generally it holds $\mathbf{Z}_J \ne \mathbf{M}_J$ and $\mathbf{D} \ne \mathbf{M}_J^{-1}\mathbf{M}$. If $\mathbf{D}$ differs not too much from $\mathbf{1}$ (times a constant) then one can obtain the multiplicands iteratively:

$$\mathbf{D}_0 = \mathbf{M} \qquad \mathbf{D}_{q+1} = \mathbf{N}(\mathbf{D}_q^{-1})\mathbf{D}_q \quad (\text{or } \mathbf{D}_{q+1} = (\mathbf{N}(\mathbf{D}_q))^{-1}\mathbf{D}_q)$$
$$\mathbf{D} = \lim_{q \to \infty} \mathbf{D}_q \qquad (\mathbf{D} := (\det \mathbf{D})^{-1/4}\mathbf{D}) \qquad \mathbf{Z}_J = \mathbf{M}\mathbf{D}^{-1}$$
$$(34)$$

$\mathbf{D}$ initially equals $\mathbf{M}$. Then its nondepolarizing content is successively split off until $\mathbf{D}$ is purely depolarizing. Indeed, after a few iterations over $q$ this converges with machine accuracy. The first expression for $\mathbf{D}_{q+1}$ can be replaced by the second, in parentheses. That also works (and may look more logical) but on average it seems to converge less fast. Without the statement $\mathbf{D} := \ldots$ in parentheses, $D$ generally differs a bit from 1 (and makes $\mathbf{Z}_J$ share its amplitude DOF with $\mathbf{D}$) whereas $\mathbf{N}(\mathbf{D}^{-1}) = \mathbf{1}$. If one normalizes $\mathbf{D}$ using its determinant (i.e. if $\mathbf{D} := \ldots$ is also executed) then $\mathbf{N}(\mathbf{D}) = \mathbf{1}$ and $\mathbf{N}(\mathbf{D}^{-1}) = D \cdot \mathbf{1}$ is obtained.

Factorization with opposite order $\hat{\mathbf{M}} = \hat{\mathbf{D}}\hat{\mathbf{Z}}_J$ is likewise possible. For instance we can set $\mathbf{M} = \hat{\mathbf{M}}^{-1}$, factorize it exactly as above, and then obtain $\hat{\mathbf{Z}}_J = \mathbf{Z}_J^{-1}$, $\hat{\mathbf{D}} = \mathbf{D}^{-1}$. Or else we factorize

$$\mathbf{M} = \mathbf{D}\mathbf{Z}_J \qquad \mathbf{N}(\mathbf{D}) = D \cdot \mathbf{1} \qquad \mathbf{N}(\mathbf{Z}_J) = \mathbf{Z}_J, \qquad (35)$$

$$\mathbf{D}_0 = \mathbf{M} \qquad \mathbf{D}_{q+1} = \mathbf{D}_q \mathbf{N}(\mathbf{D}_q^{-1}) \quad (\text{or } \mathbf{D}_{q+1} = \mathbf{D}_q (\mathbf{N}(\mathbf{D}_q))^{-1})$$
$$\mathbf{D} = \lim_{q \to \infty} \mathbf{D}_q \qquad (\mathbf{D} := (\det \mathbf{D})^{-1/4}\mathbf{D}) \qquad \mathbf{Z}_J = \mathbf{D}^{-1}\mathbf{M}$$
$$(36)$$

The sum of depolarizer matrices is also purely depolarizing,

$$\mathbf{N}(\mathbf{D}_A) = D_A \cdot \mathbf{1} \qquad \mathbf{N}(\mathbf{D}_B) = D_B \cdot \mathbf{1}$$
$$\Rightarrow \quad \mathbf{N}(\mathbf{D}_A + \mathbf{D}_B) = (D_A + D_B) \cdot \mathbf{1}. \qquad (37)$$

Any Mueller-Jones matrix can be decomposed into a product of a symmetric partial polarizer matrix $\mathbf{P}$ (23) and a retarder matrix $\mathbf{R}$ (21). In the case $\mathbf{M}_J = \mathbf{PR}$ the leftmost column of $\mathbf{P}$ equals that of $\mathbf{M}_J$. From its elements the whole $\mathbf{P}$ (23) can be calculated, and then also $\mathbf{R}$. In the case $\mathbf{M}_J = \mathbf{RP}$ the topmost row of $\mathbf{P}$ equals that of $\mathbf{M}_J$, and so on.

IV. MUELLER MATRIX MEASUREMENT, ADVANCED POLARIMETER CALIBRATION

Mueller matrix measurement is straightforward [6]: A polarization scrambler or transformer generates a sequence of $n \ge 4$ test polarizations which are measured as a reference (REF) and arranged in a $4 \times n$ matrix

$$\mathbf{S}_{REF} = [\mathbf{S}_1, \mathbf{S}_2, \ldots, \mathbf{S}_n]. \qquad (38)$$

Polarizations $\mathbf{S}_{REF}$ must span a body of non-zero volume in the Poincaré sphere. Advantageous is a low condition number of $\mathbf{S}_{REF}$. The lowest possible, $\sqrt{3}$, is obtained for tetrahedron, diamond, cube, certain other polyhedrons and equidistributed polarizations.

Then the device under test (DUT) is inserted. The resulting test polarizations are measured and arranged in a matrix

$$\mathbf{S}_{DUT} = \mathbf{M}\mathbf{S}_{REF}. \qquad (39)$$

The Mueller matrix is obtained by inversion, for instance

$$\mathbf{M} = \left(\mathbf{S}_{DUT}\mathbf{S}_{REF}^T\right)\left(\mathbf{S}_{REF}\mathbf{S}_{REF}^T\right)^{-1}. \qquad (40)$$

If appropriate, $\mathbf{M}_J$ is condensed from $\mathbf{M}$.

Measuring high PDL is not a matter of having a polarimeter with a low input power range. To understand this consider an ideal polarizer. No matter what the input polarization is, it will always output the same constant polarization, with an intensity which depends on input polarization. But the worse the polarizer extinction is, the more will the output polarization change when intensity drops to, say, half its maximum. So, to measure high PDL it suffices to measure polarization accurately at high and medium (and not necessarily at ultralow) intensities.

The insertion of DUT may consist in applying control signals to a device which was already in the path when $\mathbf{S}_{REF}$ was measured, or opening connectors and inserting the DUT there, or using a 2×2 switch which establishes a through connection for the measurement of $\mathbf{S}_{REF}$ and passes the signal through the DUT when $\mathbf{S}_{DUT}$ is measured. Drift of the polarizations generated by the scrambler/transformer and its PDL are canceled in (40) provided that between the measurements $\mathbf{S}_{REF}$, $\mathbf{S}_{DUT}$ not too much time has elapsed.

Clearly $\mathbf{S}_{REF}$ measurements which are repeated to overcome drift become cumbersome if each time the reference path must be established manually (by taking out the DUT). So, usage of the 2×2 switch is very advantageous. But the path change may falsify $\mathbf{S}_{DUT}$ measurement. Fortunately it is possible to overcome polarization drift as well as loss and PDL of the switch without establishing manual paths repeatedly:

See Fig. 1 and (41). The 2×2 switch is available to switch between paths $R$ (formerly *REF*) and $D$ (formerly *DUT*). The actual reference, for instance a through connection with a patchcord A, is inserted in a measurement $D0$. In order to assure identical test polarizations this is done right after or before a measurement $R0$. Using the switch, $\mathbf{S}_{R0}$ and $\mathbf{S}_{D0}$ are measured. Once this is done, polarizations may drift or may be changed. The true DUT is inserted in a measurement $D1$. Right before or after this, measurement $R1$ is made with the same test polarizations. $\mathbf{S}_{R1}$ and $\mathbf{S}_{D1}$ are measured in switch positions $R$ and $D$, respectively. This way the true DUT ($D1$) can be measured against the actual reference ($D0$). Polarization drift is compensated and repeated manual reference measurements are not needed. The scheme shall be presented in detail together with arising depolarization, PDL and retarders.

Let us assume polarization states $\mathbf{S}_I$ are available with constant power; subscript $I$ stands for input. They drift over time or are generated anew so that we have them in two different (matrix) versions $\mathbf{S}_{I0}$, $\mathbf{S}_{I1}$. Even their number can be different in the two versions. To suggest this their individual Stokes vectors $\mathbf{S}_{I0,i}$, $\mathbf{S}_{I1,j}$ carry different indexes $i, j$. In the measurements, $\mathbf{P}_{RS}$ describes the PDL caused by the scrambler and the switch input of path $R$. $\mathbf{R}_R$ stands for an unknown retarder in the path $R$. $\mathbf{P}_{RP}$ is the PDL at the switch output of path $R$ and in the polarimeter. $\mathbf{D}$ describes the depolarization in the polarimeter. $\mathbf{P}_{DS}$, $\mathbf{P}_{DP}$ are like $\mathbf{P}_{RS}$, $\mathbf{P}_{RP}$ but in the path $D$.

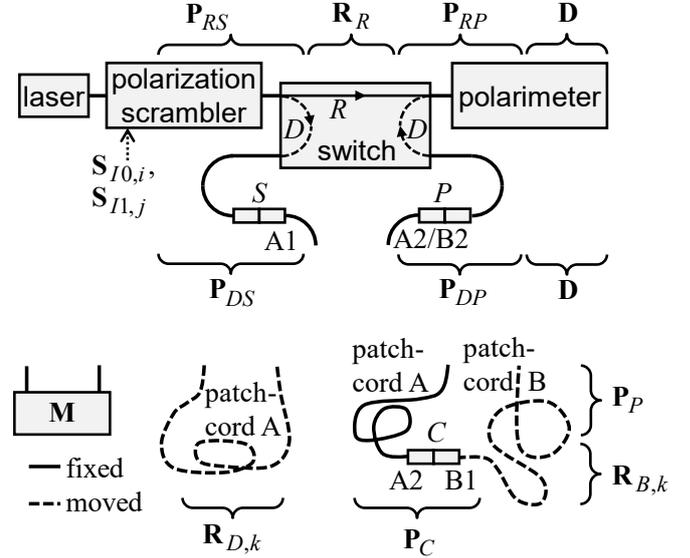

Fig. 1: Measurement setup (top) with three possible devices-under-test (bottom), and Mueller matrices describing various sections. M = general. Patchcord A with connectors A1, A2 = for calibration. Added patchcord B with connectors B1, B2 = more calibration and measurement of PDL in connection $C$. Connections (sleeves) $S$, $P$, $C$ are of same type.

Measurement $\mathbf{S}_{D0,i,k}$ is indexed according to $i = 1...n$ input polarizations and $k = 1...r$ measurements, each of them with a different position or bending of a patchcord A in path $D$, each of which is described by a different retarder $\mathbf{R}_{D,k}$. Without loss of generality we can assign to this retarder a unity matrix in those measurements $\mathbf{S}_{D0,i}$ which are kept among the many $\mathbf{S}_{D0,i,k}$, or could be taken separately, for the later calculation of $\mathbf{M}$ which describes the DUT:

$$\begin{aligned}
\mathbf{S}_{R0,i} &= \mathbf{D}\mathbf{P}_{RP}\mathbf{R}_R\mathbf{P}_{RS}\mathbf{S}_{I0,i} \\
\mathbf{S}_{D0,i} &= \mathbf{D}\mathbf{P}_{DP}\mathbf{P}_{DS}\mathbf{S}_{I0,i} \\
\mathbf{S}_{D0,i,k} &= \mathbf{D}\mathbf{P}_{DP}\mathbf{R}_{D,k}\mathbf{P}_{DS}\mathbf{S}_{I0,i} \qquad (41)\\
\mathbf{S}_{R1,j} &= \mathbf{D}\mathbf{P}_{RP}\mathbf{R}_R\mathbf{P}_{RS}\mathbf{S}_{I1,j} \\
\mathbf{S}_{D1,j} &= \mathbf{D}\mathbf{P}_{DP}\mathbf{M}\mathbf{P}_{DS}\mathbf{S}_{I1,j} \qquad (42)
\end{aligned}$$

Initially $\mathbf{D}$, $\mathbf{P}_{DP}$ must be found, using $\mathbf{S}_{D0,i,k}$. This is done like described by (16)–(19). But the $n$ scrambler polarization states arrive with different powers due to $\mathbf{P}_{DS}$. Therefore for each $i$ one sets $\mathbf{I} = \mathbf{S}_{D0,i,k}$ and starts with $\mathbf{F}'_0 = \mathbf{1}$. Each obtained end result is factorized as $\mathbf{F}'_0 = \mathbf{D}_k\mathbf{Z}_{J,k}$. The various $\mathbf{D}_k$ are averaged and give the total $\mathbf{D}$. According to (37) this is correct. The various $\mathbf{Z}_{J,k}$ are decomposed into

R. Noe, B. Koch, 28.06.2022

polarizer times retarder as described above, or one simply takes the leftmost column of $\mathbf{Z}_{J,k}$ as the leftmost column $\mathbf{P}_{0,k}$ of the polarizer matrix. The various $\mathbf{P}_{0,k}$ are averaged. Then $\mathbf{P}_{DP}$ is determined using (23). We usally set $T_{ga}=1$. Now $\mathbf{D}^{-1}$ and $\mathbf{P}_{DP}^{-1}$ can be calculated.

Evaluation of $\mathbf{S}_{D0}$ against $\mathbf{S}_{R0}$ yields

$$\begin{aligned}\left(\mathbf{S}_{D0}\mathbf{S}_{R0}^T\right)&=\mathbf{DP}_{DP}\mathbf{P}_{DS}\left(\mathbf{S}_{I0}\mathbf{S}_{I0}^T\right)\mathbf{P}_{RS}^T\mathbf{R}_R^T\mathbf{P}_{RP}^T\mathbf{D}^T\\ \left(\mathbf{S}_{R0}\mathbf{S}_{R0}^T\right)&=\mathbf{DP}_{RP}\mathbf{R}_R\mathbf{P}_{RS}\left(\mathbf{S}_{I0}\mathbf{S}_{I0}^T\right)\mathbf{P}_{RS}^T\mathbf{R}_R^T\mathbf{P}_{RP}^T\mathbf{D}^T\\ \mathbf{M}_{DR0}&=\left(\mathbf{S}_{D0}\mathbf{S}_{R0}^T\right)\left(\mathbf{S}_{R0}\mathbf{S}_{R0}^T\right)^{-1}\\ &=\mathbf{DP}_{DP}\mathbf{P}_{DS}\mathbf{P}_{RS}^{-1}\mathbf{R}_R^T\mathbf{P}_{RP}^{-1}\mathbf{D}^{-1}\end{aligned} \quad (43)$$

This is kept for future use.

Evaluation of $\mathbf{S}_{D1}$ (42) against $\mathbf{S}_{R1}$ yields

$$\begin{aligned}\left(\mathbf{S}_{D1}\mathbf{S}_{R1}^T\right)&=\mathbf{DP}_{DP}\mathbf{MP}_{DS}\left(\mathbf{S}_{I1}\mathbf{S}_{I1}^T\right)\mathbf{P}_{RS}^T\mathbf{R}_R^T\mathbf{P}_{RP}^T\mathbf{D}^T\\ \left(\mathbf{S}_{R1}\mathbf{S}_{R1}^T\right)&=\mathbf{DP}_{RP}\mathbf{R}_R\mathbf{P}_{RS}\left(\mathbf{S}_{I1}\mathbf{S}_{I1}^T\right)\mathbf{P}_{RS}^T\mathbf{R}_R^T\mathbf{P}_{RP}^T\mathbf{D}^T\\ \mathbf{M}_{DR1}&=\left(\mathbf{S}_{D1}\mathbf{S}_{R1}^T\right)\left(\mathbf{S}_{R1}\mathbf{S}_{R1}^T\right)^{-1}\\ &=\mathbf{DP}_{DP}\mathbf{MP}_{DS}\mathbf{P}_{RS}^{-1}\mathbf{R}_R^T\mathbf{P}_{RP}^{-1}\mathbf{D}^{-1}\end{aligned} \quad (44)$$

Using available results we calculate

$$\mathbf{M}_{DR1}\mathbf{M}_{DR0}^{-1}=\mathbf{DP}_{DP}\mathbf{MP}_{DP}^{-1}\mathbf{D}^{-1} \quad (45)$$

and finally the desired

$$\mathbf{M}=\mathbf{P}_{DP}^{-1}\mathbf{D}^{-1}\mathbf{M}_{DR1}\mathbf{M}_{DR0}^{-1}\mathbf{DP}_{DP}. \quad (46)$$

V. CALIBRATION AND MEASUREMENT OF CONNECTOR PDL

Using (43)–(46) and $\mathbf{M}_J=\mathbf{N}(\mathbf{M})$ we can now accurately measure even very small PDL such as connector PDL. But this holds only if: Fiber (patchcord A) in path $D$ (which was moved/bent while measuring the $\mathbf{S}_{D0,i,k}$) is cut. Then the cut ends are connectorized and connected and their PDL is measured.

Clearly this is not how connector PDL is typically assessed. We therefore expand the foregoing.

Connector PDL can only be measured jointly for two pigtails connected together, not for one pigtail alone. Let us assume the DUT in path $D$, represented by $\mathbf{M}$ in (46), was a patchcord A with input connector A1 at port $S$ (scrambler side) and output connector A2 at port $P$ (polarimeter side). It is described by a pure retarder matrix.

We write $\mathbf{P}_{DP}=\mathbf{P}_{0P}\mathbf{P}_{A2}\mathbf{R}_{02}$. $\mathbf{P}_{0P}$ describes the PDL from connector $P$ to polarimeter. $\mathbf{P}_{A2}$ describes PDL of connector A2. Retarder matrix $\mathbf{R}_{02}$ makes $\mathbf{P}_{DP}$ symmetric [5]. If the polarizations with extreme transmissions are the same in $\mathbf{P}_{0P}$ and $\mathbf{P}_{A2}$ then $\mathbf{R}_{02}$ equals $\mathbf{1}$. These polarizations might be different if, say, narrow-key FC/APC connectors are screwed into a wide-key sleeve, with unwanted angle twist. $\mathbf{R}_{02}$ does not matter here.

Connection $P$ is opened and another patchcord B is inserted there, with input connector B1 and output connector B2. Connection $C$ is where connectors A2 and B1 mate. The retarder matrix of patchcord A, measured as $\mathbf{M}$ in (46), but now in a new position or bending status, is $\mathbf{R}_0$. It holds

$$\mathbf{S}_{S1,j,k}=\mathbf{DP}_{0P}\underbrace{\mathbf{P}_{B2}\mathbf{R}_{B2}\mathbf{R}_{B,k}\mathbf{R}_{B1}\mathbf{P}_{B1}}_{\text{patchcord B}}\mathbf{P}_{A2}\mathbf{R}_{02}\mathbf{R}_0\mathbf{P}_{DS}\mathbf{S}_{I1,j}$$

(47)

where $\mathbf{P}_{B1}$, $\mathbf{P}_{B2}$ are symmetric partial polarizer matrices describing connectors B1, B2 and $\mathbf{R}_{B1}$, $\mathbf{R}_{B2}$ are retarder matrices. Now $\mathbf{D}$ and the PDL of connection $P$ and polarimeter could be determined while varying $\mathbf{R}_{B,k}$. But $\mathbf{D}$ probably has been determined before, with higher accuracy due to many different polarization settings. Also $\mathbf{P}_{DP}$ and its inverse $\mathbf{P}_{DP}^{-1}=\mathbf{R}_{02}^{-1}\mathbf{P}_{A2}^{-1}\mathbf{P}_{0P}^{-1}$ are known from that. So it makes sense to calculate

$$\begin{aligned}\mathbf{S}_{S2,j,k}&=\mathbf{P}_{DP}^{-1}\mathbf{D}^{-1}\mathbf{S}_{S1,j,k}=\mathbf{R}_{02}^{-1}\mathbf{P}_{A2}^{-1}\mathbf{P}_{0P}^{-1}\mathbf{D}^{-1}\mathbf{S}_{S1,j,k}\\ &=\underbrace{\mathbf{R}_{02}^{-1}\mathbf{P}_{A2}^{-1}\mathbf{P}_{B2}\mathbf{R}_{B2}}_{\mathbf{P}_P}\mathbf{R}_{B,k}\underbrace{\mathbf{R}_{B1}\mathbf{P}_{B1}\mathbf{P}_{A2}\mathbf{R}_{02}\mathbf{R}_0}_{\mathbf{P}_C}\mathbf{P}_{DS}\mathbf{S}_{I1,j}.\end{aligned} \quad (48)$$

$\mathbf{P}_P$ is the <u>added</u> PDL at connection $P$, by taking away the original connector A2 and inserting connector B2. $\mathbf{R}_{B2}$ may be freely chosen, for instance such that $\mathbf{P}_P$ is symmetric.

Like in (22)–(26), for each $j$ we calculate $\mathbf{P}_{P,0,j}^{-1}$, i.e. the leftmost column of $\mathbf{P}_P^{-1}$ when the scrambler produces polarization $j$. The $\mathbf{P}_{P,0,j}^{-1}$ are averaged and give a $\mathbf{P}_{P,0}^{-1}$ from which $\mathbf{P}_P^{-1}$ is calculated. It makes sense to set its $T_{ga}$ equal to 1, given that the expectation value of mean loss in $\mathbf{P}_P$ is 0 dB.

Like (42), we rearrange (47) as

$$\mathbf{S}_{S1,j,k}=\mathbf{DP}_{DP}\underbrace{\mathbf{P}_P\mathbf{R}_{B,k}\mathbf{P}_C}_{\mathbf{M}_k}\mathbf{P}_{DS}\mathbf{S}_{I1,j}. \quad (49)$$

Using (43)–(46) we can determine the various $\mathbf{M}_{J,k}=\mathbf{N}(\mathbf{M}_k)$. We calculate all $\mathbf{P}_P^{-1}\mathbf{M}_{J,k}=\mathbf{R}_{B,k}\mathbf{P}_{C,k}$. Without measurement errors, all $\mathbf{P}_{C,k}$ would be identical, i.e. a single $\mathbf{P}_C$. Each $\mathbf{P}_{C,k}$ can be determined because its topmost row is the topmost row of $\mathbf{P}_P^{-1}\mathbf{M}_{J,k}$. Determining $\mathbf{P}_C$ from the average of these topmost rows is expected to be more accurate than if one arbitrary $\mathbf{P}_{C,k}$ is taken as $\mathbf{P}_C$. $\mathbf{R}_{B1}$ may be freely chosen, for instance such that $\mathbf{P}_C$ is symmetric.

If connectors B2 and A2 behave identically then PDL (and mean loss) of $\mathbf{P}_P$ are zero. In that case, moving or bending of patchcord B is not needed and measurement of only 1 set of $\mathbf{S}_{S1,j}$ suffices! For non-trivial $\mathbf{P}_P$ its (probably quite small) PDL vector is added [5] to that of $\mathbf{P}_C$ (unless it is compensated by fiber moving/bending as described).

## VI. EXPERIMENTAL RESULTS

The setup of Fig. 1 was realized with a 20 Mrad/s polarization scrambler and a polarimeter having 100 MHz sampling frequency and 64 MStates memory.

In order to assess accuracy we have inserted patchcord A as the DUT. It was laid out and bent in $r>100$ different positions, with the aim of producing essentially all possible retarders. Using (46) and $\mathbf{M}_J = \mathbf{N}(\mathbf{M})$ the associated PDL was determined. Depending on total measurement time (of one Mueller matrix and PDL) and number $n$ of test polarizations generated by the scrambler, accuracy limits were found (Table 1). The 4 states are tetrahedron corners. The 92 geodesic dome states are the 60 corners of a ball (truncated icosahedron) and, normalized, the centers of its 32 faces.

| Total time | PDL of patchcord A | Number $n$ of test polarizations |
|---|---|---|
| 5 μs | <0.03 dB | 4 |
| 100 μs | <0.01 dB | 4 |
| 10 ms | <0.005 dB | 92 |
| 100 ms | <0.004 dB | 92 |

Table 1: PDL measurement accuracy (rounded times)

PDL of connection $C$, caused by insertion of patchcord B, was also assessed with 92 test polarizations measured in 100 ms. Connections $C$ (connectors A2 and B1) and $S$, $P$ of same type were FC/APC, FC/UPC, LC/APC, LC/UPC or E2000. Results are summarized in Table 2. The several meausurements of one connection type mean that patchcords B or adapters $C$ (sleeves) were exchanged.

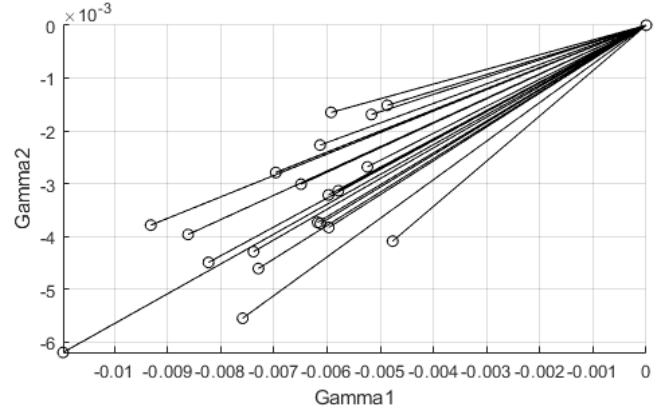
Fig. 3: PDL vectors [dB] for various positions of patchcord B

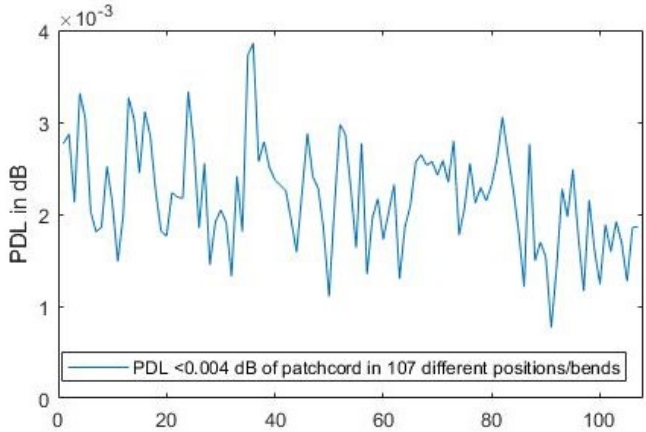
Fig. 2: Measured PDL of patchcord A with different layouts/bends

All PDL results in the case of 100 ms total measurement time for 92 test polarizations are given in Fig. 2. Mean measured PDL is about 0.0025 dB. The maximum PDL is <0.004 dB.

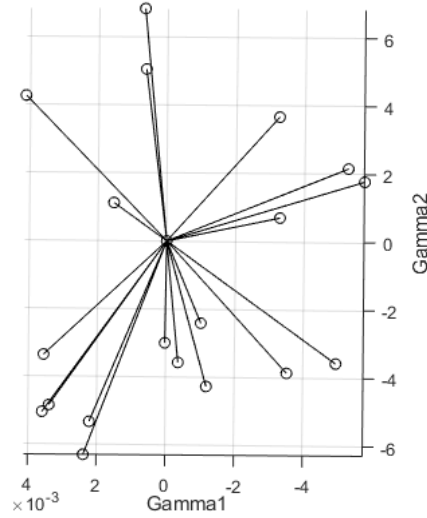
Fig. 4: PDL vectors [dB] for various positions of patchcord A

| Type of connection $C$ | PDL for various positions of patchcord B | | PDL for various positions of patchcord A | | PDL for fixed positions of patchcords A, B (repeatability) | |
|---|---|---|---|---|---|---|
| | Mean | Standard deviation | Mean | Standard deviation | Mean | Standard deviation |
| FC/APC | 0.0054 dB | 0.0016 dB | 0.0055 dB | 0.0020 dB | 0.0043 dB | 0.0006 dB |
| | 0.0079 dB | 0.0018 dB | 0.0058 dB | 0.0016 dB | 0.0054 dB | 0.0005 dB |
| | 0.0079 dB | 0.0018 dB | 0.0060 dB | 0.0009 dB | 0.0061 dB | 0.0003 dB |
| FC/UPC | 0.0113 dB | 0.0012 dB | 0.0092 dB | 0.0021 dB | 0.0083 dB | 0.0004 dB |
| | **0.0221** dB | 0.0013 dB | **0.0224** dB | 0.0032 dB | **0.0248** dB | 0.0004 dB |
| | 0.0052 dB | 0.0017 dB | 0.0064 dB | 0.0015 dB | 0.0063 dB | 0.0005 dB |
| LC/APC | 0.0056 dB | 0.0022 dB | 0.0070 dB | 0.0016 dB | 0.0038 dB | 0.0005 dB |
| | 0.0047 dB | 0.0009 dB | 0.0042 dB | 0.0009 dB | (not measured) | |
| LC/UPC | **0.0201** dB | 0.0015 dB | **0.0170** dB | 0.0020 dB | **0.0153** dB | 0.0009 dB |
| | 0.0056 dB | 0.0021 dB | 0.0076 dB | 0.0023 dB | 0.0050 dB | 0.0006 dB |
| | 0.0085 dB | 0.0014 dB | 0.0082 dB | 0.0019 dB | 0.0061 dB | 0.0004 dB |
| E2000 | 0.0062 dB | 0.0012 dB | 0.0078 dB | 0.0019 dB | 0.0086 dB | 0.0004 dB |
| | 0.0068 dB | 0.0012 dB | 0.0071 dB | 0.0026 dB | 0.0071 dB | 0.0007 dB |

Table 2: PDL measurement of various connections. Results >0.015 dB are boldface.

PDL in dB is $|\mathbf{\Gamma}_{dB}|$ (24), the length of the dB-scaled extinction-based PDL vector. In Figs. 3–5 the extinction-based PDL vectors in dB $\mathbf{\Gamma}_{dB}$ are plotted for an exemplary case (last line FC/APC, underlined). Stokes parameter axes are multiplied by dB. When only patchcord B is moved the PDL vector does not change very much (Fig. 3). When only patchcord A is moved, the direction of the PDL vector changes very much, unlike its length (Fig. 4). When patchcords are fixed the PDL vector is most stable and one simply checks repeatability in vectorial form (Fig. 5).

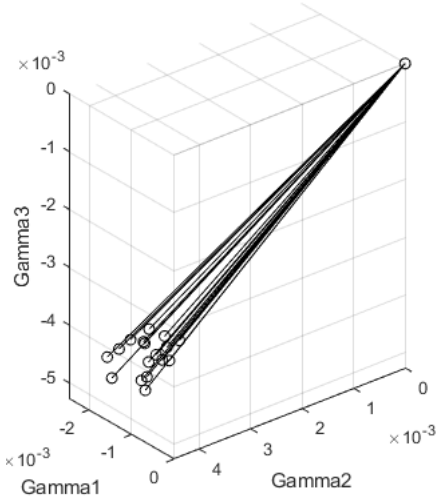

Fig. 5: PDL vectors [dB] for fixed positions of patchcords A, B (repeatability)

## VII. Discussion

In a couple of cases, connector PDL was >0.015 dB (boldface in Table 2). This looked random, or reasons are unknown. But in earlier tests with various FC/APC adapters, involving mixture of wide-key and narrow-key, we got PDL from about 0.01 dB up to 0.06 dB. In particular, PDL changed when the body of the connector was turned left while the nut was screwed and fastened to the right. In one case PDL was 0.0511 dB and 0.0528 dB for two different positions of patchcord A. As expected, PDL vector direction changed from 1st to 2nd measurement and the small difference 0.0017 dB between these two PDL values is in line with the PDL measured for patchcord A alone. Also there, moving only patchcord B kept length and direction of the PDL vector fairly constant.

Generally, high connector PDL seems to be related to high mean insertion loss (maybe sometimes caused by dirt).

The accurate results show the validity of the theory and prove that connector PDL can be measured with low error <0.004 dB.

For the testing of multiple connectors one can easily place 1:$m$ and $m$:1 switches at ports $S$ and $P$. Each of the paths will need a separate calibration.

Cascades of more inserted connectors can be characterized similarly as described above: Moving fiber(s) allows isolating PDL devices and characterizing individual device PDL accurately.

At the high end of the range we have measured PDL >60 dB of a good polarizer [7]. More precisely, PDL ranged from 62 to >80 dB, with a mean of 69 dB. PDL is usually much higher than the specified extinction ratio that also comprises PMF misalignment.

## VIII. Conclusion

After suitable polarimeter calibration, polarization-dependent loss from <0.004 dB to >60 dB was measured. This is made possible by a 20 Mrad/s polarization scrambler with LiNbO$_3$ polarization transformer and a polarimeter with 100 MHz sampling frequency and 64 MStates memory. Connectors with lowest PDL and polarizers with extreme extinction are successfully characterized.